\documentclass[preprint,showpacs,preprintnumbers,amsmath,amssymb,aps]{revtex4}

\usepackage{graphicx}
\usepackage{dcolumn}
\usepackage{bm}
\usepackage{natbib}
\usepackage{braket}
\usepackage{units}

\renewcommand{\d}{ \text{d}}
\newcommand{\X}{X^{++}}
\newcommand{\be}{\begin{equation}}
\newcommand{\ee}{\end{equation}}

\begin{document}

\preprint{DO-TH-11/03}

\title{Neutrino production of hadrons \\at low energy and in the small $Q^2$ region}

\author{E. A. Paschos}
 \email{paschos@physik.uni-dortmund.de}
\author{Dario Schalla}
 \email{dario.schalla@tu-dortmund.de}
\affiliation{Department of Physics, TU Dortmund, D-44221 Dortmund, Germany}

\date{April 19, 2011}

\begin{abstract}
We use PCAC in the small $Q^2$ region in order to calculate the Adler sum rule and the production of hadrons in the low energy region where resonances dominate. We find very good agreement with the sum rule and with the computed cross sections. We find a value $C_5^A(0)$ close to the Goldberger-Treiman prediction. The formalism is general and can be applied to other reactions shedding light into the dynamical transition from resonances to deep inelastic scattering.
\end{abstract}


\maketitle

\section{Introduction}
\label{sec:Introduction}
Inspite of the long history of neutrino interactions with hadrons there is still interest on them because many cross sections, especially at low energies, are not known precisely. Analyses of experiments need the cross sections in order to interpret properties in the oscillations of neutrinos and separate specific quantities, like the mixing angle $\theta_{13}$, $CP$-violation, etc. The cross sections are also of interest on their own right because they explain properties of the strong interactions and the transition from the resonance to the deep inelastic region.

In this article we consider again the production of hadrons in the small $Q^2$-region. Matrix elements of the vector current have been evaluated using the conserved vector current (CVC) property by relating them to electroproduction data. This way the magnitude and $Q^2$ dependence have been determined~\cite{Lalakulich:2006sw} and have been applied~\cite{Paschos:2003qr,Hernandez:2007qq,Lalakulich:2010ss} successfully.
For the vector current we use the results from~\cite{Lalakulich:2006sw}. The conservation of the vector current is already incorporated in the equations of the Appendix of~\cite{Lalakulich:2005cs}.

The matrix elements of the axial current are more difficult to determine and one resorted to models for estimating specific values of the form factors and their $Q^2$ dependence. In addition data in the small $Q^2$ region do not agree with some theoretical predictions \cite{Paschos:2003qr,Furuno:2003ng}. In this article we address the small $Q^2$ region. We adapt the principle that amplitudes which are free of strong interaction singularities in a specific variable will generally vary smoothly with that variable \cite{Weinberg1968}. We apply this smoothness assumption to the delta resonance by using the partially conserved axial vector current (PCAC) for $Q^2\leq 5 m_\pi^2 \approx \unit[0.1]{GeV^2}$ and extend the results to values twice as big, i. e. $Q^2$ up to $\unit[0.2]{GeV^2}$. This kinematic region is especially sensitive in charged current reactions because the three quantities $Q^2$, $m_\pi^2$ and $m_\mu^2$ are of the same magnitude. As a check we calculate with the same assumptions the Adler sum rule and establish that it is very well satisfied for $Q^2 \leq \unit[0.2]{GeV^2}$. For the differential cross section $\d\sigma / \d Q^2$ we use exact kinematics by keeping the three small quantities mentioned above.

We outline now the main framework and the equations that we shall use in our calculations. The matrix element of $\mathcal{A}_\mu^+$ between a proton and the delta resonance is defined in Eq.~(2.7) of reference \cite{Lalakulich:2005cs} to which we will refer as (LP). It involves the form factors $C_3^A (Q^2),\; C_4^A(Q^2),\; C_5^{A}(Q^2)$ and $C_6^A (Q^2)$. The last form factor is the induced pseudoscalar, which can be written explicitly in terms of a pion pole. When we keep the mass of the muon, the leptonic current can be expanded in four polarization vectors \cite{Paschos:2005km,Paschos:2009ag}. Estimates of contributions from transverse polarizations are small in the small $Q^2$ region and will be neglected. The reason for this is the small $Ap\rightarrow \pi p$ amplitude which is primarily longitudinal~\cite{Ballam:1968sw}. Thus we shall concentrate on contributions from the polarizations
\be
\epsilon_s^\mu = \frac{q^\mu}{\sqrt{Q^2}}
\label{eq:polS}\ee
and
\be
\epsilon_\mu (\lambda=0) = \frac{1}{\sqrt{Q^2}} \left( |\vec{q}| ,\; 0,\; 0,\; q_0 \right).
\label{eq:pol0}\ee

For the small $Q^2$ region we adopt these approximations. In this case the overall tensors multiplying the form factors $C_3^A$ and $C_4^A$ are such that their inner product with $q_\mu$ vanishes (see Eq. (2.7) in reference\cite{Lalakulich:2005cs}).

We keep the two form factors $C_5^A$ and $C_6^A$ and for the pseudoscalar form factor we write the pion pole explicitly. For the $\pi p \Delta$ vertex we introduce $\Braket{\Delta | j_\pi | p}$ with $j_\pi$ the pion source:
\be
\Braket{\Delta^{++} | \mathcal{A}_\mu^+ | p} = \sqrt{3} \; \overline{\psi}_\lambda (p') i g^\lambda_\mu C_5^A (Q^2) u (p) + \sqrt{3} \frac{i f_\pi q_\mu}{q^2-m_\pi^2} \Braket{\Delta | j_\pi | p}.
\ee
Taking the divergence on both sides and using PCAC we arrive at
\begin{eqnarray}
i  q^\mu \Braket{\Delta^{++} | \mathcal{A}_\mu^+ | p} &=& \sqrt{3} \; \overline{\psi}_\lambda (p') q^\lambda C_5^A (Q^2) u(p) + \sqrt{3} \frac{f_\pi q^2}{q^2-m_\pi^2} \Braket{\Delta | j_\pi | p} \label{eq:PPA} \\
&=& \sqrt{3} \frac{f_\pi m_\pi^2}{q^2-m_\pi^2} \Braket{\Delta | j_\pi | p}.\label{eq:PPB}
\end{eqnarray}
This relation holds for values of $Q^2$ where PCAC is valid. It follows now that
\be
C_5^A (Q^2) \overline{\psi}_\lambda (p') q^\lambda u(p) = - f_\pi \Braket{\Delta | j_\pi | p}.
\label{eq:dominant}\ee
This is our PCAC relation which we assume to hold in the extended region of $Q^2$ because there are no singularities in $Q^2$. Obviously, we eliminated the pion pole by subtracting Eq. (\ref{eq:PPB}) from (\ref{eq:PPA}).

A similar relation holds when we replace the delta resonance by a general final state $X^{++}$. In this cases
\be
\Braket{\X | \mathcal{A}_\mu^+ | p} = \mathcal{R}_\mu + \sqrt{3} \frac{i f_\pi q_\mu}{q^2-m_\pi^2} \Braket{\X | j_\pi | p}
\ee
whose divergence gives the relation
\be
q^\mu \mathcal{R}_\mu = - i f_\pi \Braket{\X | j_\pi | p}.
\ee
Here $\mathcal{R}_\mu$ is the remaining amplitude beyond the pion pole.

The reader should note that this relation holds for any final state $m_X \neq m_p$ and also when we replace the proton in the target by a neutron. We need this relation for estimating the reaction $A(W^+n\rightarrow X^+) = A(W^-p\rightarrow X^0)$ in the sum rule and also for contributions up to $W\leq \unit[1.6]{GeV}$.

Another remarkable property of neutrino interactions is that for small values of $Q^2$ relative to the neutrino energy $E_\nu$ and energy transfer $\nu$ the leptonic current can be approximated with the two polarizations given in Eq. (\ref{eq:polS}) and (\ref{eq:pol0}). The scalar polarization gives a small term proportional to $m_\pi^2$ which we keep. The polarization with zero helicity annihilates the pion pole and gives the dominant contribution for the remainder as follows:
\begin{eqnarray} \epsilon^\mu (\lambda =0) \Braket{\Delta^{++} | \mathcal{A}_\mu | p} &=& \epsilon^\mu (\lambda =0) i \sqrt{3} \overline{\psi}_\mu (p') C_5^A (Q^2) u(p) \\
& \approx & i \sqrt{3} \frac{q^\mu}{\sqrt{Q^2}} \overline{\psi}_\mu (p') C_5^A (Q^2) u (p) + \mathcal{O} \left( \frac{Q^2}{\nu^2}  \right) \\
& \approx & - \frac{f_\pi \sqrt{2}}{\sqrt{Q^2}} A (\pi^+ p \rightarrow \Delta^{++}) \label{eq:eA}
\end{eqnarray}
with $f_\pi = \unit[0.093]{GeV}$.
We emphasize that we first eliminate the pion pole and then approximate the helicity zero polarization  by $q_\mu / \sqrt{Q^2}$. This method leaves out terms of $\mathcal{O} \left( Q^2 / \nu^2 \right)$ and justifies the application of PCAC. 
Thus the polarization with zero helicity takes the divergence of the axial current.
Using this result we calculate in the following sections the differential cross section and the Adler sum rule as a function of $Q^2$.

In previous articles the axial contribution was accounted for by
introducing form factors and estimating the $Q^2$ dependence from neutrino data. The Goldberger-Treiman relation that was obtained earlier \cite{Lalakulich:2006sw,Schreiner:1973mj} is also inherent in Eq. (\ref{eq:dominant}) when we substitute bare couplings for the matrix elements. What is different now is the replacement of the left-hand side of Eq. (\ref{eq:dominant}) by $\pi^+ p \rightarrow \Delta^{++}$ or $\pi^+ p \rightarrow \X$ data. As it will become clear later on, PCAC with the smoothness assumption provides an estimate of the form factor for $Q^2 \leq \unit[0.2]{GeV^2}$. The handling of the vector-axial interference (the structure function $W_3 (Q^2,\nu)$) will be explained in section 3.

In an extension of the model some authors \cite{Hernandez:2007qq,Lalakulich:2010ss}, in order to account for the non-resonant background, use additional diagrams generated by an effective Lagrangian. In this article we compute the contribution of the axial current using PCAC and for the vector contribution we use the results from (LP).
The production of pions through the exchange of mesons and Regge trajectories in the t-channel has been also worked out~\cite{Gershtein:1980vd}. Our PCAC results in Eqn.~(\ref{eq:dominant})-(\ref{eq:eA}) sum up the contributions of these exchanges.

\section{General formulas}
\label{sec:GeneratlFormulas}
From the results of the previous section we determine the contribution of the axial current alone to the cross section
\be
\frac{\d \sigma^A}{\d Q^2\d \nu} = \frac{G_F^2 |V_{ud}|^2}{2\pi} \frac{1}{4\pi} \frac{\nu}{E_\nu^2} \frac{f_\pi^2}{Q^2} \left\{ \tilde{L}_{00} + 2 \tilde{L}_{l0} \frac{m_\pi^2}{Q^2+m_\pi^2} + \tilde{L}_{ll} \left( \frac{m_\pi^2}{Q^2+m_\pi^2} \right)^2 \right\} \sigma (\pi^+ p \rightarrow \X).
\label{eq:axial}\ee
The matrix elements $\tilde{L}_{00}$, $\tilde{L}_{l0}$, $\tilde{L}_{ll}$ were introduced and elaborated in references \cite{Paschos:2005km,Paschos:2009ag}. The elements $\tilde{L}_{l0}$ and $\tilde{L}_{ll}$ are proportional to $m_\mu^2$ and with an additional factor of $m_\pi^2$ or $m_\pi^4$ become too small and will be neglected. The cross section $\sigma (\pi^+ p \rightarrow \X)$ is the production of hadrons at the energy $\nu$. It is evident now that integrating over $\nu$ in order to obtain $\d\sigma / \d Q^2$ we trace the delta resonance peak and this influences the turning over of the cross section at low $Q^2$.

For evaluating the Adler sum rule we need the structure function $W_2 (Q^2,\nu)$. The identification is more evident when we consider the dominant contribution in our Eq. (\ref{eq:axial}) in the limit $E_\nu - \nu \gg \nu, \; \frac{Q^2}{M},\; \frac{m_\mu^2}{M}$. The leptonic density matrix element squared was computed in \cite{Paschos:2005km} as
\be
\tilde{L}_{00} = 4 \left[ \frac{[Q^2(2E_\nu - \nu)-\nu m_\mu^2]^2}{Q^2 (Q^2+\nu^2)} - Q^2 - m_ \mu^2 \right]
\ee
which in the high energy limit becomes
\be
\tilde{L}_{00} \rightarrow \frac{2 Q^2}{Q^2 +\nu^2} \left[ 4 E_\nu E' - (Q^2 + m_\mu^2) - \frac{m_\mu^2}{Q^2} \nu^2 \right].
\ee
By comparing our Eq.~(\ref{eq:axial}) with the $W_2 (Q^2,\nu)$ in Eq. (2.10) of (LP) we find the relation
\be
W_2^A (Q^2,\nu) = \frac{2 f_\pi^2}{\pi} \frac{\nu}{Q^2 + \nu^2} \sigma (\pi^+ p \rightarrow \X)
\label{eq:W2Apcac}\ee
with the superscipt $A$ denoting the axial contribution alone.

\section{Testing the Adler sum rule}
\label{sec:AdlerSumRule}
A basic relation in particle physics is the Adler sum rule for neutrinos \cite{Adler:1968hc}. It follows from current algebra and holds for each value of $Q^2$. For the axial current it takes the form
\be
\left[ g_A (Q^2) \right]^2 + \int_{\nu_{th}}^\infty \d \nu \left[ W_{2,\nu n}^A (Q^2,\nu ) - W_{2,\nu p}^A (Q^2,\nu ) \right] = 1
\ee
with $g_A(Q^2)$ the form factor for the vertex $\Braket{p|\mathcal{A}^+_\mu|n}$.

With the functional form for $W_2^A (Q^2,\nu)$ given in Eq. (\ref{eq:W2Apcac}) we obtain
\be
\left[ g_A (Q^2) \right]^2 + \frac{2 f_\pi^2}{\pi} \int_{\nu_{th}}^\infty \d \nu \frac{\nu}{Q^2 + \nu^2} \left[ \sigma^{\pi^-p} (\nu) - \sigma^{\pi^+ p} (\nu) \right] = 1 .
\label{eq:sumrule}\ee
The functional form in Eq. (\ref{eq:sumrule}) follows from the general formalism. It is a very convenient form because at $Q^2 = 0$ it reduces to the Adler-Weisberger relation \cite{Adler:1965ty,Weisberger:1966ip}.

We tested the relation by using experimental data for the pion-nucleon scattering and
\be
g_A (Q^2) = \frac{-1.26}{\left( 1 + \frac{Q^2}{M_A^2} \right)^2}
\ee
with $M_A=\unit[1.0]{GeV}$. 

The data we use are from the particle data group \cite{Nakamura:2010zzi} and from Barashenkov et al.~\cite{Barashenkov}. Both sources give compilations from a large number of experiments. The data gives both the elastic and total cross sections for the reations $\pi^+ p$ and $\pi^- p$ which is equal to $\pi^+ n$. The values of the cross section did not change in the last few decades. In the region of integration the inelastic channel for $\pi^+ p$ is negligible. For $\pi^- p$ it is substantial which indicates the presence of non-resonant background. For the sum rule we use the total cross sections. In the next section we shall use data for the $\pi^+ p$ channel for extracting $C_5^A$, where as mentioned the difference between elastic and total cross section is negligible.

In order to give some typical value we tabulated in Tab.~\ref{tab:h0010} and \ref{tab:h0150} values of the cross sections for two different values of $Q^2= \unit[0.010]{GeV^2}$ and $Q^2 = \unit[0.150]{GeV^2}$. The cross sections correspond to the values of $W$. In the last column we give values for the variable $A=\frac{\nu f_\pi^2 \tilde{L}_{00}}{E_\nu^2 Q^2}$ which enters Eq.~(\ref{eq:axial}) as an overall factor. The cross section is the product of $A$ with the pion-nucleon cross section which indicates that the overwhelming contribution us close to the delta resonance peak.

\begin{figure}
\includegraphics{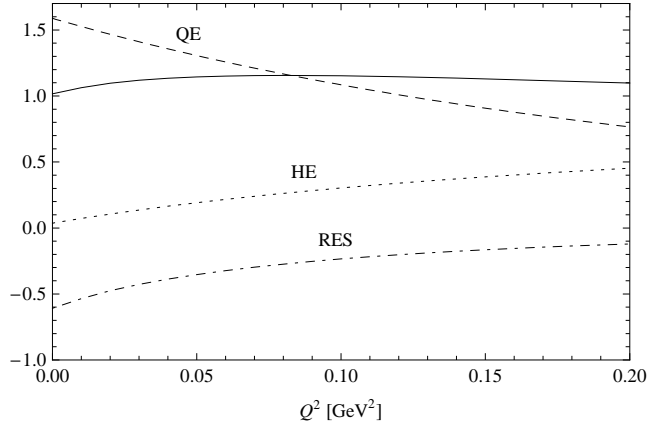}
\caption{Solid: Adler sum; dashed: form factor; dot-dashed: integral over resonances; dotted: high energy part (Bodek-Yang interpolation).}
\label{fig:sumrule}
\end{figure}

The results of the numerical integration are shown in Fig.~\ref{fig:sumrule} with the resonance contribution corresponding to the integration of the total cross sections. The integral has been truncated at $\nu = \unit[1.6]{GeV}$. The sum of QE and RES should saturate the sum rule. The difference 1.0 - (QE+RES) corresponds to contributions from higher energies, where multipion production is important.
This is indicated in the figure as HE. 
The HE contribution is very small for $Q^2 \sim 0.0$ and increases monotonically indicating that the multipion contribution is more important for larger values of $Q^2$. Bodek and Yang, after analyzing data in deep inelastic scattering~\cite{BodekYang}, studied the transition to the resonance region and suggested an interpolation, which agrees rather well with our estimated value from the sum rule. The sum of the three contributions produces the value of one to 10~\% or better. The errors in the hadronic cross sections are less than 3~\%. Thus the confirmation of the sum rule is an indication for the validity of PCAC.

\begin{table}
\begin{tabular}{|c|c|c|c|c|}
\hline
$\nu$ [GeV]	&	$W$ [GeV]	&	$\sigma (\pi^+ p)$ [mb]	&	$\sigma (\pi^- p)$ [mb]	&	$\frac{\nu f_\pi^2 \tilde{L}_{00}}{E_\nu Q^2}$ [GeV] \\
\hline
0.20	&	1.118	&	16	&	12	&	0.329	\\
0.25	&	1.159	&	77	&	30	&	0.240	\\
0.30	&	1.199	&	189	&	67	&	0.171	\\
0.35	&	1.238	&	175	&	63	&	0.119	\\
0.40	&	1.275	&	95	&	37	&	0.079	\\
0.45	&	1.311	&	60	&	28	&	0.047	\\
0.50	&	1.347	&	42	&	26	&	0.022	\\
0.55	&	1.381	&	31	&	28	&	0.002	\\
\hline
\end{tabular}
\caption{Values for $E_\nu = \unit[1]{GeV}$ and $Q^2 = \unit[0.010]{GeV}$.}
\label{tab:h0010}
\end{table} 

\begin{table}
\begin{tabular}{|c|c|c|c|c|}
\hline
$\nu$ [GeV]	&	$W$ [GeV]	&	$\sigma (\pi^+ p)$ [mb]	&	$\sigma (\pi^- p)$ [mb]	&	$\frac{\nu f_\pi^2 \tilde{L}_{00}}{E_\nu Q^2}$ [GeV] \\
\hline
0.20	&	1.053	&	16	&	12	&	0.109	\\
0.25	&	1.097	&	77	&	30	&	0.113	\\
0.30	&	1.139	&	189	&	67	&	0.111	\\
0.35	&	1.180	&	175	&	63	&	0.104	\\
0.40	&	1.219	&	95	&	37	&	0.095	\\
0.45	&	1.257	&	60	&	28	&	0.085	\\
0.50	&	1.294	&	42	&	26	&	0.074	\\
0.55	&	1.330	&	31	&	28	&	0.063	\\
0.60	&	1.364	&	24	&	28	&	0.052	\\
0.65	&	1.398	&	17	&	34	&	0.042	\\
\hline
\end{tabular}
\caption{Values for $E_\nu = \unit[1]{GeV}$ and $Q^2 = \unit[0.150]{GeV}$.}
\label{tab:h0150}
\end{table}

\section{The differential cross section}
\label{sec:differentialcrosssection}
For the calculation of the various terms we apply different methods because we choose for each current the method that is most reliable. For the axial current we use the PCAC result in Eq.~(\ref{eq:axial}). For the vector current we use the form factors from reference~\cite{Lalakulich:2006sw}. The vector-axial interference term is obtained by using formula (A3) in LP with the vector form factors from~\cite{Lalakulich:2006sw,Lalakulich:2005cs} and the axial form factor $C_5^A (Q^2)$ extracted later on in this section. We elaborate on each of these terms.

\begin{enumerate}
\item
For the axial contribution we use the data for pion-nucleon cross sections~\cite{Nakamura:2010zzi}. We concentrate on the reaction $\nu_\mu p \rightarrow \mu^- p \pi^+$ where the non-resonant background is negligible so that we select $\X = \Delta^{++}$. We integrate the cross section over $\nu$ tracing the product $\tilde{L}_{00} (Q^2,\nu,E_\nu) \sigma_{\pi p} (\nu)$ and we obtain the curve denoted as PCAC in Fig.~\ref{fig:E01000}. Our method shows that the turn over in $Q^2$ is in part a reflection of the peak in the invariant mass of the resonance. In the same figure we show the contributions from $(C_3^V)^2$, the interference term and the remaining form factors (rest) which are smaller.

We also determine the axial form factor $C_5^A (Q^2)$ by computing the integral $\int_{\nu_{th}}^{\unit[1.6]{GeV}} W_2^A (Q^2,\nu) \d \nu$ numerically using Eq.~(\ref{eq:W2Apcac}) and as a second method we use Eq. (A2) of (LP) where only $C_4^A$ and $C_5^A$ are kept. The two results contain axial-current contributions and by comparison we extract values for $C_5^A (Q^2)$ as shown in Fig.~\ref{fig:C5A}. For data we used the reaction $\pi^+ p \rightarrow \Delta^{++}$ where the non-resonant background is smallest. It is reassuring that the value $C_5^A (0) = 1.08$ is close to $1.20$ predicted by the Goldberger-Treiman relation.
In the same figure we plotted $C_5^A (Q^2)$ from the parametrization in (LP), noting that the two results are close to each other.
Several recent articles calculated $C_5^A(0)$ by fitting the experimental data~\cite{Hernandez:2007qq,Lalakulich:2010ss,Leitner:2008ue,Graczyk:2009qm,Hernandez:2010bx,AlvarezRuso:1998hi,SajjadAthar:2009rc} with their values varying from 0.87 up to 1.20. Models with a resonant background~\cite{Hernandez:2007qq,Lalakulich:2010ss} prefer the power value, while the other articles~\cite{Leitner:2008ue,Graczyk:2009qm,Hernandez:2010bx,AlvarezRuso:1998hi,SajjadAthar:2009rc} prefer values closer to 1.20. The reasons for the differences is the treatment of the non-resonant background and the exact kinematics at small $Q^2$.

\begin{figure}
\includegraphics{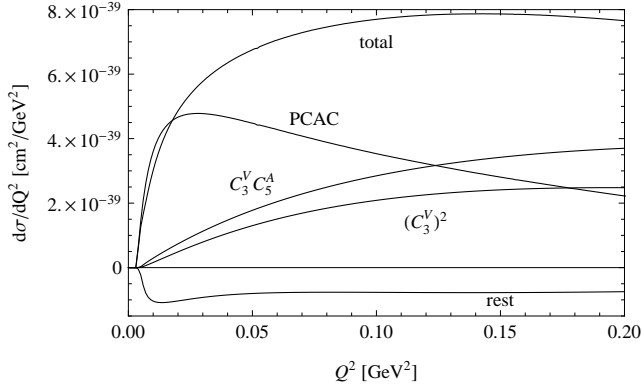}
\caption{Contributions to the total cross section at $E_\nu = \unit[1]{GeV}$ with no cuts.}
\label{fig:E01000}
\end{figure}

\begin{figure}
\includegraphics{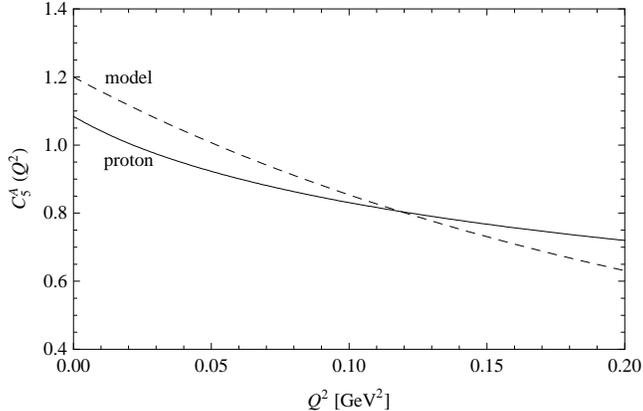}
\caption{Extracted $C_5^A$ values for proton scattering (solid) and the phenomenological model $C_5^A(Q^2)= \frac{1.2}{(1+Q^2/M_A^2)^2}\frac{1}{1+2Q^2/M_A^2}$ with $M_A=\unit[1.05]{GeV}$ (dashed). Throughout our calculation we use the extracted values.}
\label{fig:C5A}
\end{figure}

\item
For the vector contribution we use the formulas from \cite{Lalakulich:2006sw} 
\begin{eqnarray}
 C_3^V (Q^2) = \frac{2.13 / D_V}{1+ \frac{Q^2}{4M_V^2}} \; & , & \; C_4^V (Q^2) = \frac{-1.51 / D_V}{1+ \frac{Q^2}{4M_V^2}} \\
 C_5^V (Q^2) = \frac{0.48 / D_V}{1+\frac{Q^2}{0.776 M_V^2}} \; & \textmd{ and } & \; D_V =\left( 1 + \frac{Q^2}{M_V^2} \right)^2
\end{eqnarray}
with $M_V = \unit[0.84]{GeV}$, which have been extracted from electroproduction data. The result is also shown in Fig. \ref{fig:E01000}.

\item 
For the vector-axial interference $W_3 (Q^2,\nu)$ we use the form factor $C_5^A (Q^2)$ extracted through PCAC, $C_4^A= -\nicefrac{1}{4} C_5^A$ and the vector form factors just described. Then we use the formulas from (LP) and calculate their contributions to the cross section.
The three contributions are shown separately in Fig. \ref{fig:E01000}.
\end{enumerate}

\noindent
We note that the three contributions are comparable and are important for determining the $Q^2$ dependence of  $\d \sigma / \d Q^2$.

\section{Comparisons}
\label{sec:Comparisons}
\begin{figure}
\includegraphics{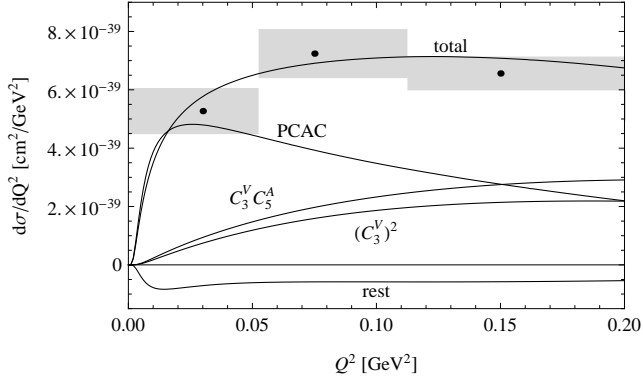}
\caption{Contributions to the total cross section for the ANL flux with $W< \unit[1.4]{GeV}$.}
\label{fig:ANL}
\end{figure}

\begin{figure}
\includegraphics{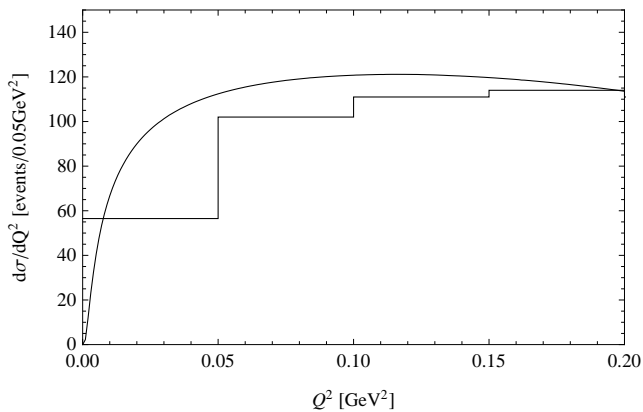}
\caption{Contributions to the total cross section for the BNL flux with $W< \unit[1.4]{GeV}$.}
\label{fig:BNL}
\end{figure}

For comparisons with data we shall account for two bubble chamber experiments~\cite{Radecky:1981fn,Kitagaki:1986ct} and the recent results from MiniBooNE \cite{AguilarArevalo:2010bm}.
The results of our calculation using the Argonne experiment (ANL) are shown in Fig.~\ref{fig:ANL}. We weighted the theoretical curve with the ANL flux and limited $W<\unit[1.4]{GeV}$. 
A similar comparison was performed for the Brookhaven (BNL) experiment shown in Fig.~\ref{fig:BNL}. The experimental data is given in terms of event rates \cite{Kitagaki:1986ct}. For comparison with absolute cross sections we use the transformation coefficient obtained in~\cite{Lalakulich:2010ss}.
In both cases the curves are close to the experimental points. We emphasize that we did not include the non-resonant background, which has been estimated in electroproduction to be 10 \% \cite{Galster:1972rh,Bartel:1968tw}.

\begin{figure}
\includegraphics{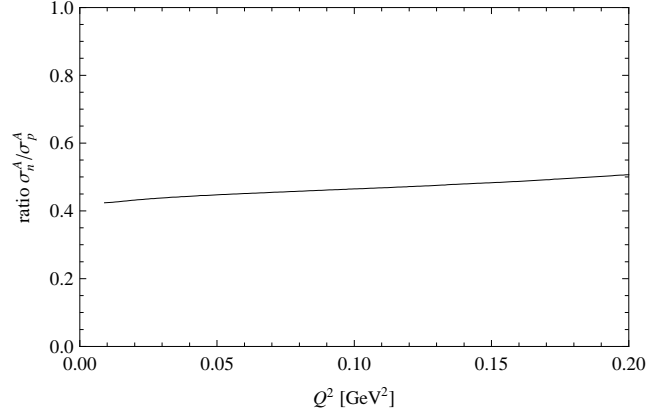}
\caption{Ratio $\sigma^A_n / \sigma^A_p$ for the MiniBooNE flux.}
\label{fig:axialratio}
\end{figure}

\begin{figure}
\includegraphics{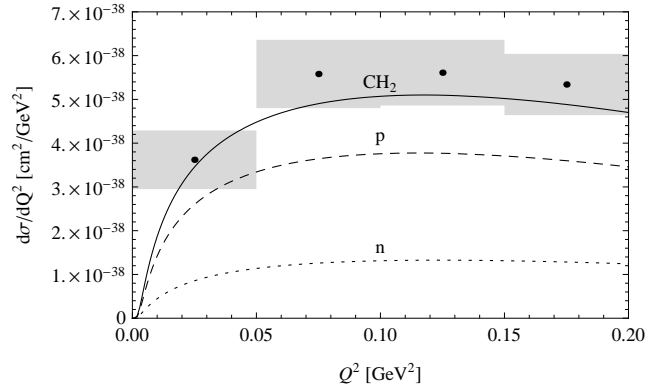}
\caption{Contributions to the total cross section for the MiniBooNE flux with $W< \unit[1.4]{GeV}$. Coherent sum of proton and neutron cross section for CH$_2$. Neutron cross section are scaled proton cross section by ratio of direct axial contribution ratio of proton and neutron data.}
\label{fig:MiniBooNE}
\end{figure}

The MiniBooNE group also reported data for a CH$_2$ target and an absolute normalization~\cite{AguilarArevalo:2010bm}. For the nuclear targets we use the incoherent sum of scatterings on neutrons and protons. We use for each channel the same formulas and the appropriate hadronic cross sections. 
For this we calculate also the cross section on neutron targets using hadronic data. For the ratio of neutrino scattering we found the ratio $\sigma_n^A/\sigma_p^A$ shown in Fig.~\ref{fig:axialratio}.
We weighted the cross sections with the neutrino spectrum and show the results in Fig.~\ref{fig:MiniBooNE} by taking 8 proton and 6 neutron cross sections. Again we did not consider explicitly the non-resonant background or any rescattering (FSI) that may occur in the carbon target. Note that the results in Fig.~\ref{fig:MiniBooNE} are for the entire molecular target which makes the cross section very big.

\begin{figure}
\includegraphics{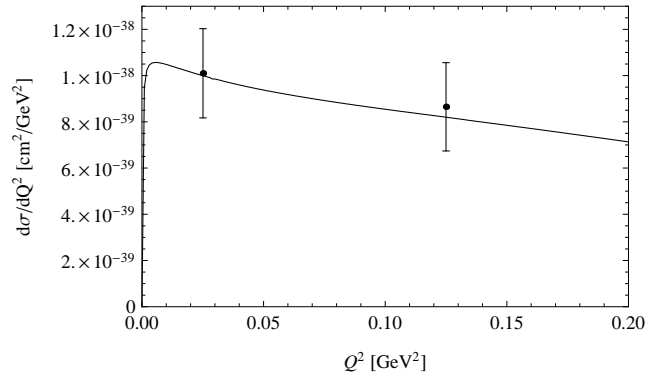}
\caption{Contributions to the total cross section for the FNAL-15ft experiment with $W< \unit[1.4]{GeV}$. Neutrino energy fixed at $\unit[25]{GeV}$.}
\label{fig:FNAL}
\end{figure}

Finally we consider a very high energy reaction from the FNAL 15ft bubble chamber~\cite{Bell:1978rb} and the results are shown in Fig.~\ref{fig:FNAL}.

\section{Summary and outlook}
\label{sec:SummaryAndOutlook}
We applied the PCAC relation to the small $Q^2$ region and calculated the contribution of the axial current. An interesting result is the expression for $W_2^A(Q^2,\nu)$ in terms of pion nucleon scattering cross sections, Eq.~(\ref{eq:W2Apcac}). We use the new expression to compute the Adler sum rule and showed that it is saturated almost completely by data in the delta energy region, leaving a smaller contribution from higher energies, which becomes more important as $Q^2$ increases. The higher energy contribution suggests an interpolation given in footnote~\cite{BodekYang}.

We also studied extensively a general formula for the differential cross section which includes the mass of the charged lepton exactly. Our estimates of the cross section for charged pion production by neutrinos shows that contributions from vector, axial and interference terms are comparable. The addition of all three determines the magnitude, position and shape of the peak in the differential cross section at $Q^2 \leq \unit[0.20]{GeV^2}$. Encouraged from the results we computed the differential cross sections for the Argonne, Brookhaven and MiniBooNE experiments. The comparisons in Fig. \ref{fig:ANL}-\ref{fig:MiniBooNE} are very good. Besides the low energies a calculation for $E_\nu = \unit[25]{GeV}$ produces the measured cross section. We did not include the non-resonant background which can be the subject for further studies \cite{Lalakulich:2010ss}. For this reason we concentrated on the $\Delta^{++}$ production where the background is the smallest. For medium heavy nuclei rescattering corrections will also be required.

Our method allows to extract the axial form factor $C_5^A(Q^2)$ whose value at $Q^2=0$ is consistent with the Goldberger-Treiman relation. This form factor has been the subject of several articles~\cite{Hernandez:2007qq,Lalakulich:2010ss,Leitner:2008ue,Graczyk:2009qm,Hernandez:2010bx,AlvarezRuso:1998hi,SajjadAthar:2009rc} with its value varying from 0.87 to 1.20. We feel that a precise treatment of the various contributions and of mass effects are essential.

Our approach can be extended and combined with experimental data in order to predict cross sections at other energies, for reactions with antineutrinos and also for neutral currents.
The kinematic region we considered is the place where coherent scattering on nuclei also occurs. Our cross section should be useful in subtracting the resonant contribution leaving as a remainder the coherent production.
The PCAC relation and the cross section in Eq.~(\ref{eq:axial}) are general and can be applied to other resonances.

Finally, the rapid saturation of the sum rule and estimates of the cross sections hint how the transition from resonances to the deep inelastic region is attained dynamically. The picture that emerges for $\d \sigma /\d Q^2$ includes a constant contribution from the algebra of commutators plus a term varying with $Q^2$ from the form factor and low mass resonances which fade away as $Q^2$ increases, leaving space for multiple final states. Our results support and extent the results obtained earlier \cite{Lalakulich:2006yn}.

\begin{acknowledgements}
We wish to thank Dr. O. Lalakulich for a critical reading of the manuscript and together with Prof. A. Bodek for private communications concerning the sum rule.
One of us (EAP) thanks the Humboldt Foundation for a travel grant to Fermilab which encouraged this research. 
\end{acknowledgements}


\begin{thebibliography}{10}
\bibitem{BodekYang}
A.~Bodek and U.~Yang,
\newblock (private communications) A convenient parametrization for the high
  energy contribution is $$\delta W_2 = 2 \frac{Q^2+0.19}{Q^2+0.29} -
  \frac{Q^2+0.25}{Q^2 +0.20}$$ which improves the sum rule. We are informed
  that work on this topic is still in progress.
\end{thebibliography}
\end{document}